# A Conglomerate of Multiple OCR Table Detection and Extraction

Smita Pallavi
Birla Institute of Technology, Patna
smita.pallavi@bitmesra.ac.in

Raj Ratn Pranesh*
Birla Institute of Technology, Mesra
raj.ratn18@gmail.com

Sumit Kumar*
Birla Institute of Technology, Mesra
sumit.atlancey@gmail.com

***Abstract* -** Information representation as tables are compact and concise method that eases searching, indexing, and storage requirements. Extracting and cloning tables from parsable documents is easier and widely used, however industry still faces challenge in detecting and extracting tables from OCR documents or images. This paper proposes an algorithm that detects and extracts multiple tables from OCR document. The algorithm uses a combination of image processing techniques, text recognition and procedural coding to identify distinct tables in same image and map the text to appropriate corresponding cell in dataframe which can be stored as Comma-separated values, Database, Excel and multiple other usable formats.

## I. Introduction

OCR documents or images of typed text and tables are a major challenge for parsing task to aid editing, searching, indexing and compact storage of documents. Table is a common and important form of representation and storage of information. Widely used as a form of documentation and information storage these promote interests of bank sectors, insurance domains, computerized receipts and various other domains. Extracting multiple tables from OCR documents and images is a widespread and challenging task because of its implementation and algorithmic complexity. Even simple tables are not indexed by systems easily. We have proposed a three stage algorithm to extract multiple tables from images and separate them using image processing, text recognition and procedural coding. This algorithm can efficiently extract all the tables from an image, digitize and separate to ease parsing.
The remainder of this paper is organized as follows:
Section 2 discusses related literature reviews and its shortcomings. Section 3 describes about Image processing method used in the algorithm presented in this paper. Section 4 describes the proposed algorithm, it's flowchart and pseudocode of the various stages. Section 5 shows the output analysis and the algorithms efficiency. Finally, we conclude with summary and further research.

## II. RELATED LITERATURE REVIEW

Various methods to detect tables and identify topological structures of images have been formulated in recent past.

*Equal Contribution

B. Freisleben et.al(2004)[1] proposed an efficient method to localize, binarize and segment text constituted in digital images. Gatos B. et.al(2005)[2] worked on table detection using horizontal and vertical lines. Several other table detection and segmentation methods in heterogeneous documents works are proposed[3-5].

Yefeng Zheng, Changsong Liu, Xiaoqing Ding and Shiyan Pan(2001)[6] proposed one of the most efficient form frame line detection algorithm which used single chain connected method. A Rehman, F Kurniawan & T Saba (2011)[7] removed the smashing of characters while line detection and removal from OCR images. Thotreingam Kasar, Philippine Barlas(2013)[8] detected table region by identifying column and row separators by applying a run-length approach to identify vertical and horizontal lines. Manolis Vasileiadis, Nikolaos Kaklanis, Konstantinos Votis, Dimitrios Tzovaras (2017)[19] proposed a method for automatic detection and extraction of Tabular data using page segmentation techniques to obtain text data and group them using bottom up technique.

In this paper[9] author R.W Smith Detect table regions from heterogeneous Document images using Layout analysis module of Tesseract on document images using Tab Stop detection. The table detection algorithm is used in identifying the table partitions, Detecting Page Column split, Locating Table Columns, Marking the table regions and Removing the false alarms.The paper present method to detect table in varying layout documents but failed to identify and spot the graphical images containing texts.

Author Cesarini et.al in paper[10] proposed a methodology to search for parallel lines of MXY tree in page. The detection is counterchecked by locating white spaces and perpendicular lines in the regions between detected parallel lines. On the basis of proximity and similarity criteria, located tables are merged. This will classify whether the images are tabular or not. This paper claims to provide table detection algorithm and index for table location evaluation, however fails to structurize the table contents

In paper[8] author Thotreingam Kasar et.al proposed a method to detect horizontal and vertical lines by run-length approach. 26 low-level features are extracted from each group of horizontal and vertical lines. SVM classifier predicts whether the it belongs to the table or not.The author used Machine Learning methods to classify table regions in heterogeneous documents without resorting to heuristic rules.

Our proposed model uses 3-stage algorithm to extract table from the image. First contours of each cell are detected using image processing techniques and grouped according to the tables. Text is extracted using Tesseract OCR engine. An algorithmic approach is used to map the text to the cell contours detected. This will detect, extract and store multiple tables into separate csv, DB, Excel files etc. Our approach uses combination of image processing and text recognition to structure table data into indexable format. This also works on presence of multiple tables in a single image.

.

## III. IMAGE PROCESSING METHODS

Advanced Image processing methods motivates cleaning, precise binarization, structural analysis and pattern extraction from images. These processing are the basic building blocks for advanced semantic analysis on images. In our table extraction algorithm we've used few binarization techniques and morphological transformations to achieve the desired output.

### A. Otsu Binarization

Otsu's binarization or thresholding method iterates through all possible values of t i.e. threshold and calculates a measure of spread for the pixel levels on both sides of the threshold value. The target is the optimization of the sum of background and foreground spreads. The optimization function which calculates the intra class variance is given by:

$$^2(\Theta) = \quad 1(\Theta) \quad 1\,^2(\Theta) + \quad 2\,(\Theta) \quad 2^2\,(\Theta) \qquad (1)$$

where Θ is the threshold, $1^2$ and $^2$ are variance of the two classes. 1 and 2 are probabilities of the two classes. 1,2(Θ) is calculated from the L bins of the histogram as given by:-

$$1(\Theta) = \sum_{i=0}^{\theta-1} P\,(i) \qquad (2)$$

$$2(\Theta) = \sum_{i=\theta}^{L-1} P\,(i) \qquad (3)$$

Otsu's Algorithm

**for each** intensity:
  compute histogram and probabilities
**end for**
  calculate initial value of $\omega_i(0)$ and $\mu_i(0)$
**for** t=1 **to** maximum intensity:
    update $\omega_i$ and $\mu_i$
    compute $\sigma_b^2(t)$
**end for**
  optimal t is at maximum $\sigma_b^2(t)$

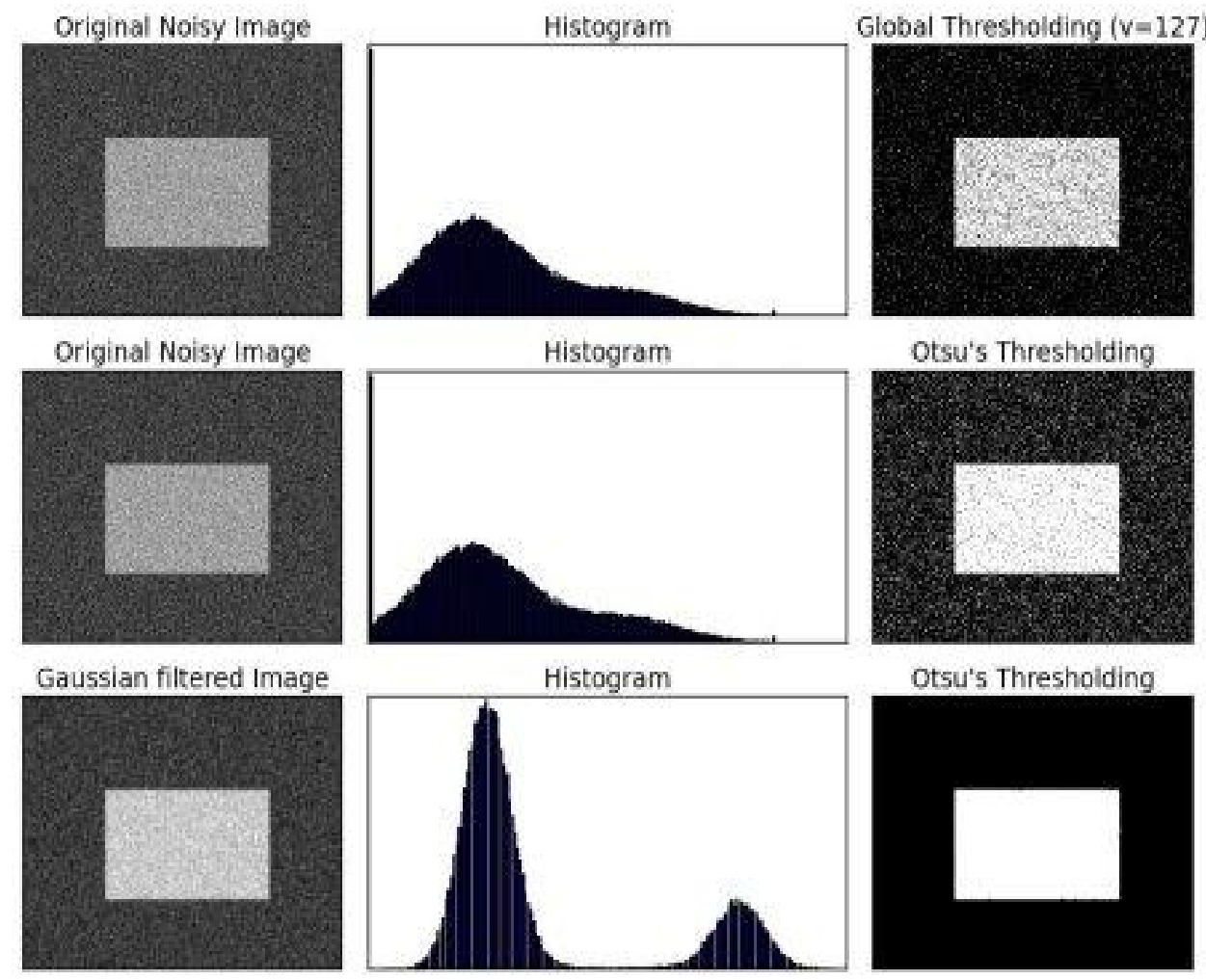

Fig -1: Result of Otsu's Thresholding with Gaussian filter

Jin Soo Noh and Kang Hyeon devised a palmprint identification [11] algorithm using Otsu binarization. It has also been used in various historic document preprocessing[12]. Adaptive Gaussian binarization[13] is used where image intensity distribution is non uniform. Adaptive binarization is an efficient way to preprocess non-uniformed images[14-15]. A fast 2D-Ostu algorithm has been devised using improved histogram[16] which shows significant improved in our uniform image preprocessing.

### B. Morphological Transformation

Image is interpreted as integer grid $Z^d$ or subset of Euclidean space $R^d$ for some dimension d in binary morphology. A binary structuring element modifies the image grid according to the fit or miss of the image grid. The structuring element that is used is a vertical and a horizontal element both of length image height by 80. This parameter is subject to experimentation on different kind of images depending upon its magnification.

For 'E' Euclidean space, Erosion of a binary image X in E using structuring element Y is:

$$X \ominus Y = \{ z \in E \mid Y_z \subseteq X \} = \bigcap_{y \in Y} X_{-y} \qquad (4)$$

where $Y_z$ is the translation of Y by the vector z. Alternatively, $X_{-y}$ denotes the translation of X by -y.

X eroded by Y can be interpreted as the locus of the points reached by the geometric center of Y when Y moves inside of X.

Dilation of binary image X by structuring element Y is denoted as:

$$X \oplus Y = \{ z \in E \mid (Y^s)_z \cap X \neq \emptyset \} = \bigcup_{y \in Y} X y \qquad (5)$$

where $Y^S$ denotes the symmetric of Y as:

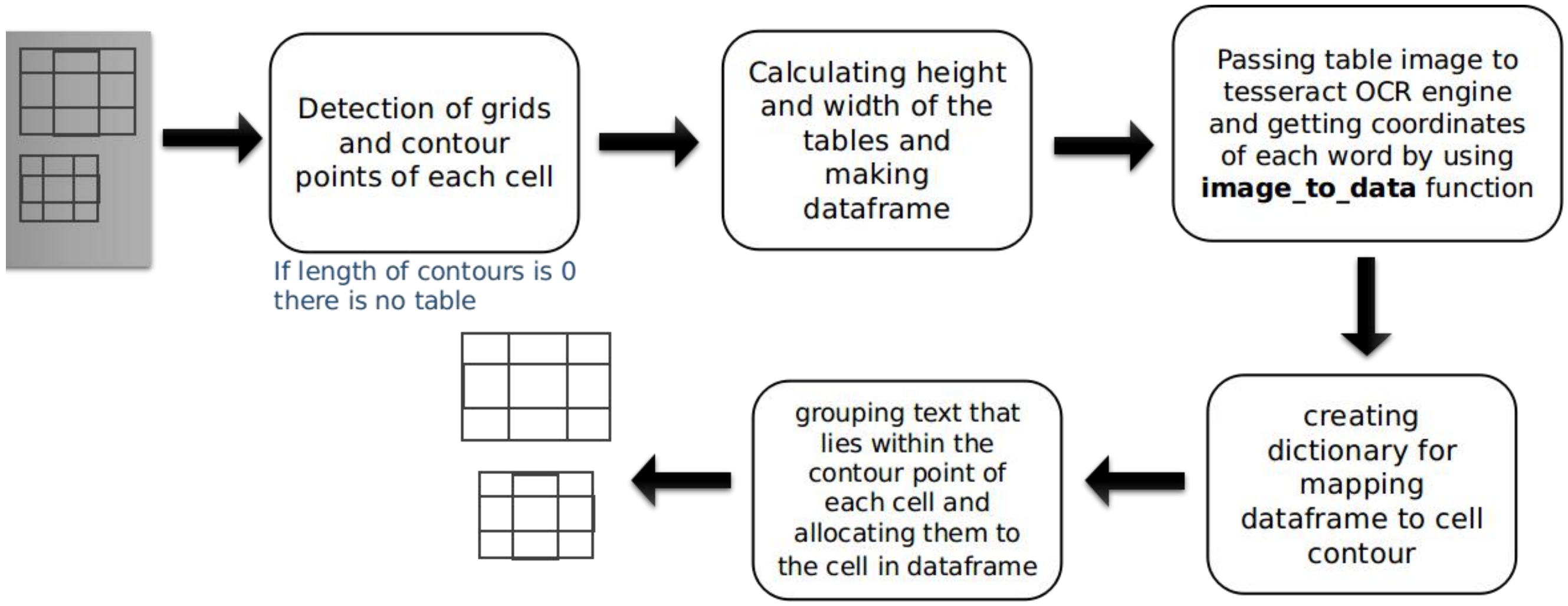

Fig. 2: Flowchart of the proposed Algorithm

$$Y^s = \{ x \in E \mid -x \in B \} \quad (6)$$

Consecutive erosion followed by dilation of a image by same structuring element is called Opening. Opening is defined as conjugation of (5) and (6) as:

$$X \circ Y = (X \ominus Y) \oplus Y = \cup_{Y_z \subseteq X} Yx \quad (7)$$

Opening can filter out the desired structure from the image. The desired structure is defined by the Structuring element. To get all the vertical lines in an image, opening is performed using vertical Structuring element. Sreedhar et.al(2012) proposed enhancements of images using morphological transformations, opening and closing[17].

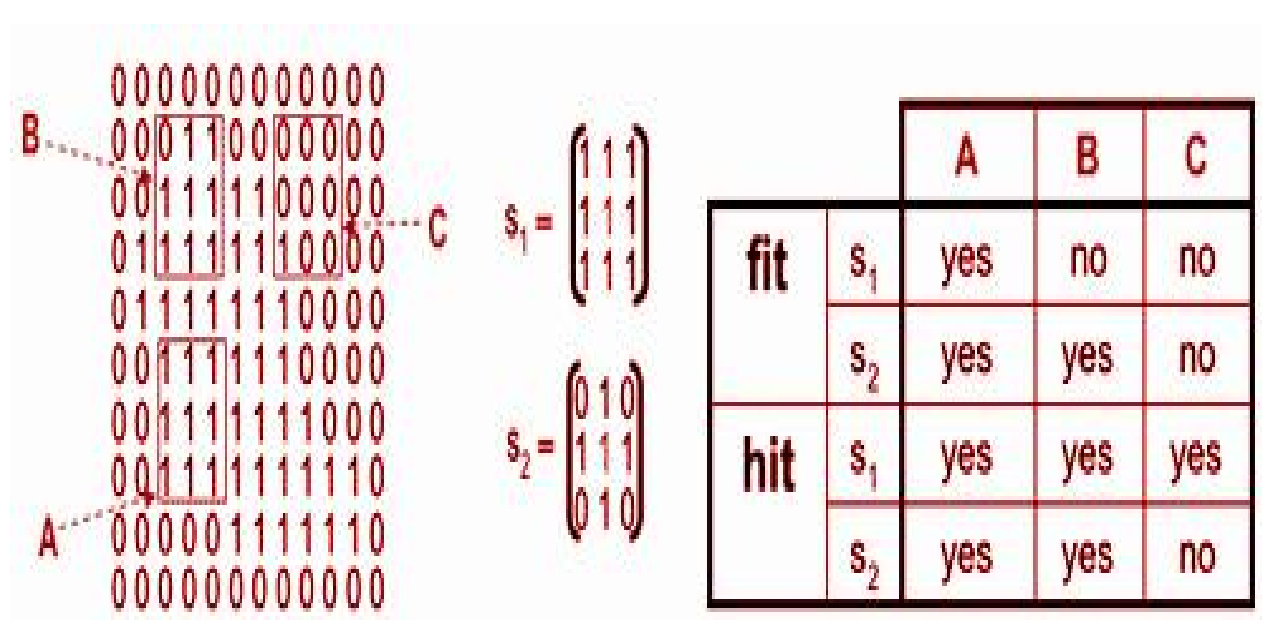

Fig -3: Fitting and hitting of an image by two structuring elements s1 and s2.

## IV. Proposed Multi-OCR extraction model

The proposed model is a 3-stage pipeline. The first stage is the application to various image processing techniques to extract the skeleton of the table. Once the skeleton is extracted we apply border following method to detect the contours of the white patches on the black background i.e. the outline of each cell of the table. Basic algorithmic approach is applied to separate different tables in the image. If the length of list of contours extracted is 0, there is no table present.

In the next stage we work on the text coordinate extraction part and storing it in such a way to easily allocate to the appropriate cell. The text is extracted using tesseract module and coordinates of each words are stored along with parameters like meanX and meanY. These two parameters helps in allocation to the appropriate cell.

Finally we apply a series of operation to integrate both the data. First, the contours are processed to detect the height and width of the table and appropriate dataframe is made store the final output. Now we create a dictionary for mapping dataframe which has key in the form of X,Y of dataframe and contour as the value of the key. Finally, we assign values to each cell of dataframe by filtering operation of the tesseract output by meanX and meanY of each word..

### *A. Stage 1: Extract table semantics*

**Table Semantics Extraction Algorithm**

input = image(converted into grayscale)

---

preprocess_image(Otsu's binarization by minimizing within class variance given by this equation)
$\sigma_\psi^2(\Theta) = \psi 1(\Theta)\, \sigma 1^2(\Theta) + \psi 2(\Theta)\, \sigma 2^2(\Theta)$

l = image_height//80
vertical_kernel, horizontal kernel, kernel = $[A]_{l,1}$, $[A]_{1,l}$, $[A]_{3,3}$ | $A_{i,j} = 1$
vertical_lines = image Θ vertical_kernel, image ⊕

```
   vertical_kernel
   horizontal_lines = image Θ horizontal_kernel, image ⊕
   horizontal_kernel
   combined_image = 0.5x vertical_image + 0.5 x
   horizontal_image
   contours = find_contours(combined_image)
while contours is not NULL:
   largest_cell = max(contours)_y
   group.append(contours_X,Y < largest_cell_X,Y)
   contours.remove(group)
end while
   If group.size is 0:
       return no table
  else:
return group
```

We start the semantics extraction by taking a grayscale image with multiple tables as input. Otsu's binarization is performed on the image by exhaustive search for *t*(threshold) that minimizes the variance within the class given the equation. Zhu, Ningbo & Wang et.al(2009) proposed a fast 2D Otsu thresholding based on improved histogram.

To extract the vertical and horizontal lines of the tables from the image we perform morphological transformation called opening which is three iteration of erosion followed by three iteration of dilation. The transformation is implemented with kernels of length page height by 80. The orientation of kernel varies for both the transformations. The contour point of each cell is extracted by border following method[18] which analyses the topological structure of the image. The four corner points are determined by combination of maximum and minimum of both x and y points of contours.

The border points extracted from images now needs to be grouped for presence of multiple table in one image. To group contour of same table we first find the contour with maximum height and group it with contours what lie within it. From the remaining group of contours we perform the same steps again and again unless the initial group is empty. Note that the find_contours function outputs a contour for the external borders of the entire table i.e. it returns both the inner and outer border for a structure.

*B. Stage 2: Getting text features from images*

We now apply tesseract OCR engine image_to_data function directly of image. The tesseract output include x, y, width, height for a word in an image. For each word output features like x_mean and y_mean are calculated. Both values make the center of the word image. The tesseract also gives confidence for each word extraction on which a threshold is applied to filter noise and unknown symbols from the image. The mean position clubbed with each word can be used to find its location within contour.

*C. Stage 3: Mapping dataframe with contours*

**Table initialization Algorithm**

```
input = image(converted into grayscale)
```

```
for table in group:
  group rows with Δcontour_y_mean < line_threshold (10px)
   make dataframe with (X=number of groups, Y=max
  element in one group) t
  row_number, column_number = 0, 0
  sort contours of table with key = y_mean, x_mean
   for i in range(len(contour)-1):
     if abs(contour[i] - contour[i+1]) < line_threshold:
        group tesseract text with x_mean and y_mean
        between contour[i] and assign to row_number,
        column_number of made dataframe
      column_number++
    else:
      row_number++
      column_number=0
         group tesseract text with x_mean and y_mean
        between contour[i] and assign to row_number,
        column_number of made dataframe
    end for
  end for
```

The final stage of extraction include mapping the derived coordinates to dataframe by filtering text from the output of tesseract. The lines' contours are groups by clubbing those with $\Delta contour_{y_mean}$ less than some threshold. By experiments we put the value as 10px. Once the groups are made, the final dataframe rows correspond to the number of groups and columns correspond to the maximum number of element in one group.

Now we iterate through the contours sorted with y-mean and then from x-mean. Whenever the difference between y-means of two consecutive contour is less than the line_threshold, text are filtered from the tesseract output whose mean lies between the contours and the column number is incremented by 1. Whenever the difference is greater than the line_threshold, row number is incremented by one and column number is reset followed by filtering of text from tesseract output.

## V. Experimental results

The image being processed is a BGR image taken from a camera. It is converted into grayscale in the first place and then the operations are carried out. The histogram of the image is plotted to analyze the intensity of distribution of the image.

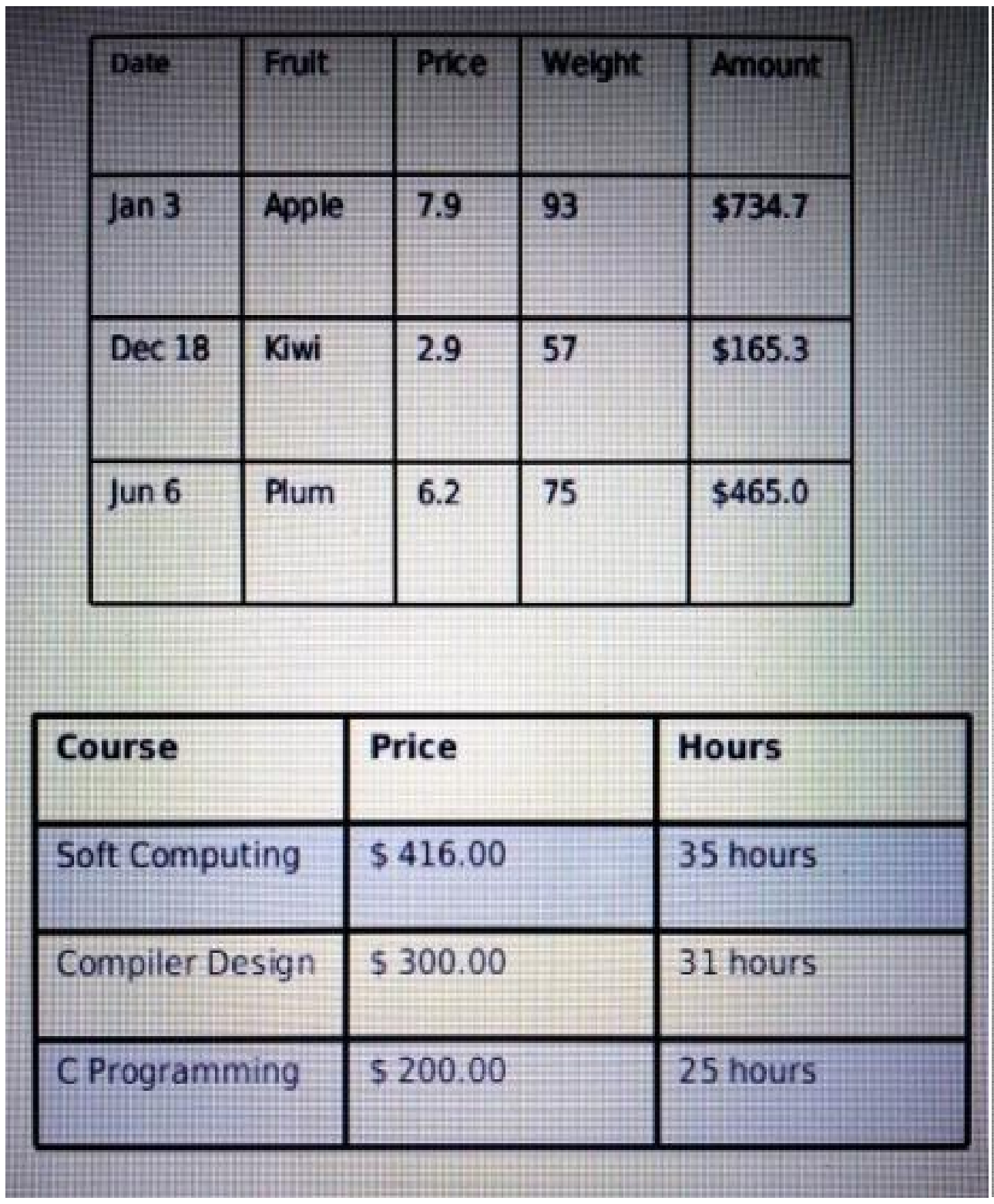

| Date | Fruit | Price | Weight | Amount |
|---|---|---|---|---|
| Jan 3 | Apple | 7.9 | 93 | $734.7 |
| Dec 18 | Kiwi | 2.9 | 57 | $165.3 |
| Jun 6 | Plum | 6.2 | 75 | $465.0 |

| Course | Price | Hours |
|---|---|---|
| Soft Computing | $ 416.00 | 35 hours |
| Compiler Design | $ 300.00 | 31 hours |
| C Programming | $ 200.00 | 25 hours |

(a)

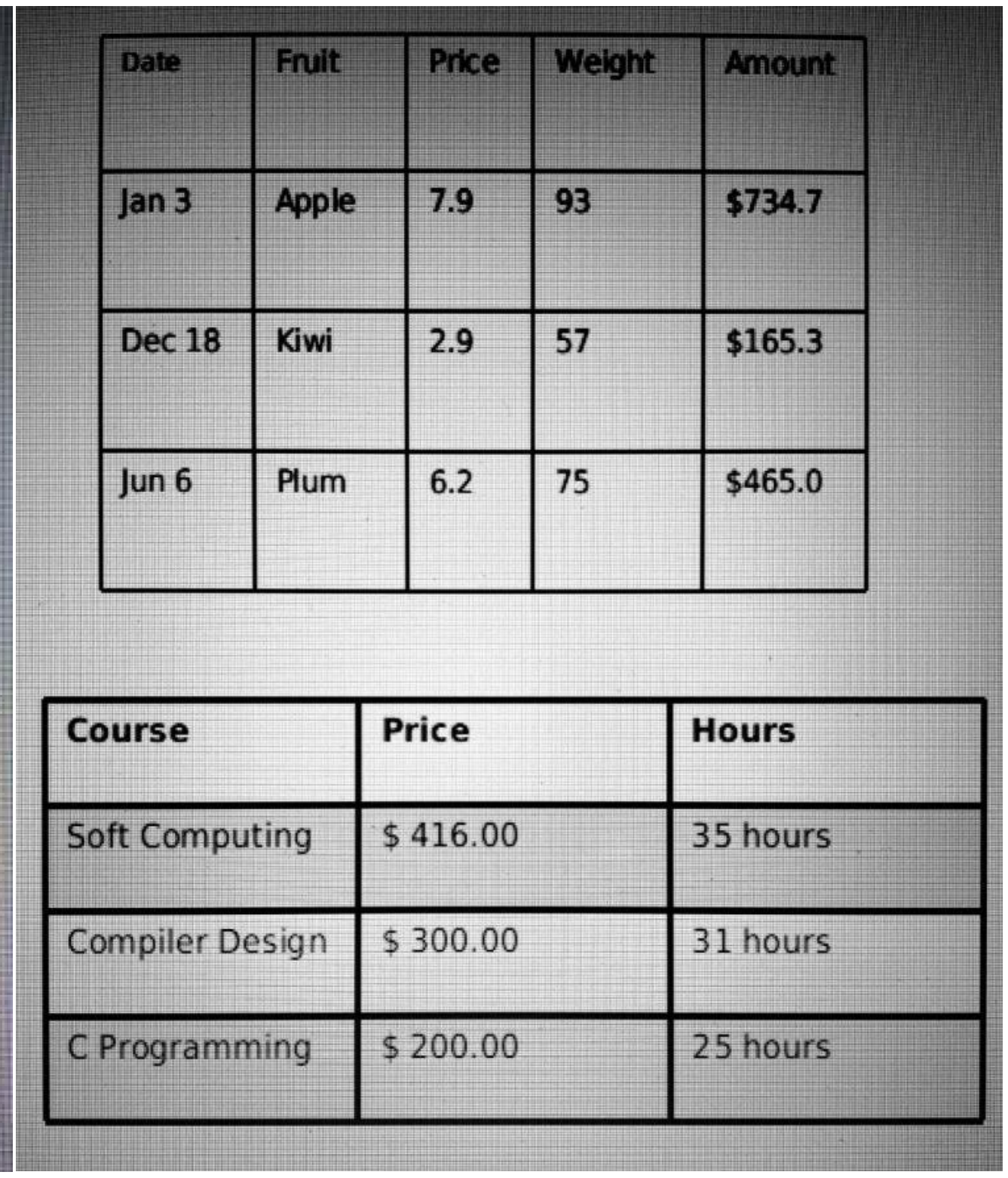

(b)

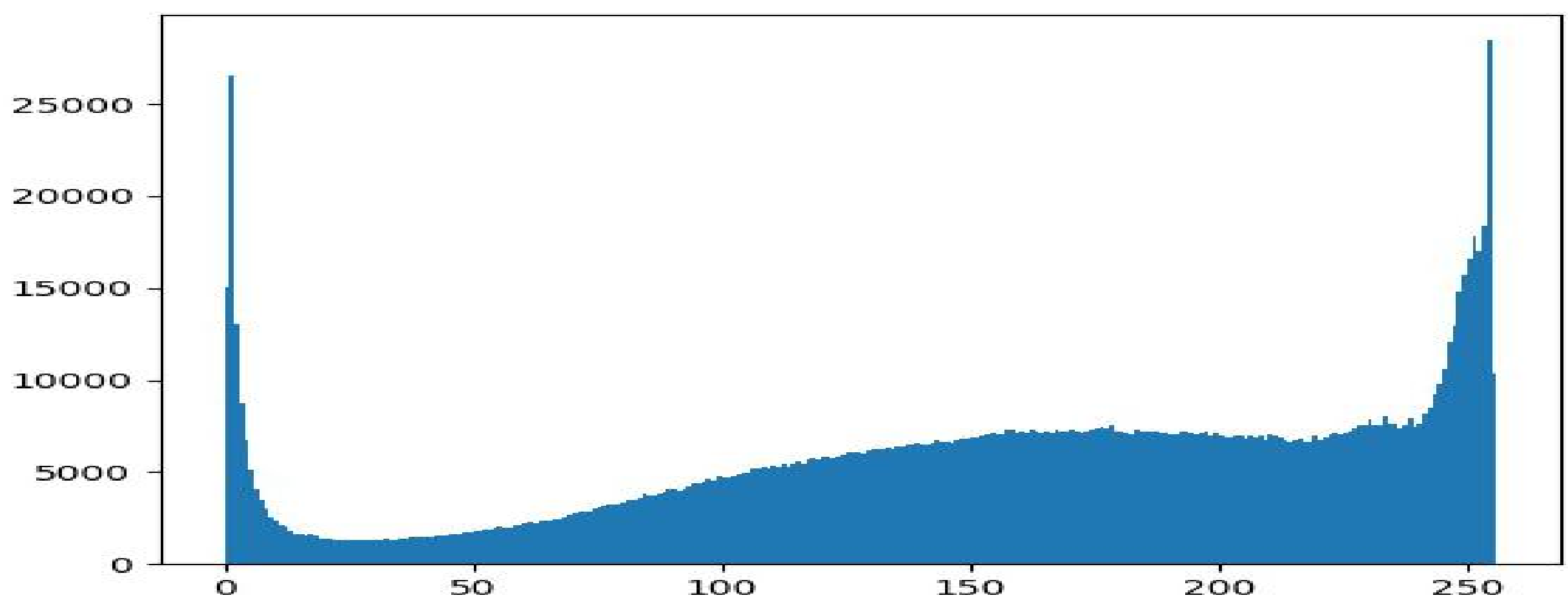

(c)

Fig -4: (a) The raw image in BGR to be processed (b) Image converted into grayscale (c) The Histogram of the image to visualize the intensity of distribution of image

We then perform threshold measurement followed by opening of the image. Otsu and Adaptive Gaussian thresholding is applied to the grayscale image to filter the boundary of the table.

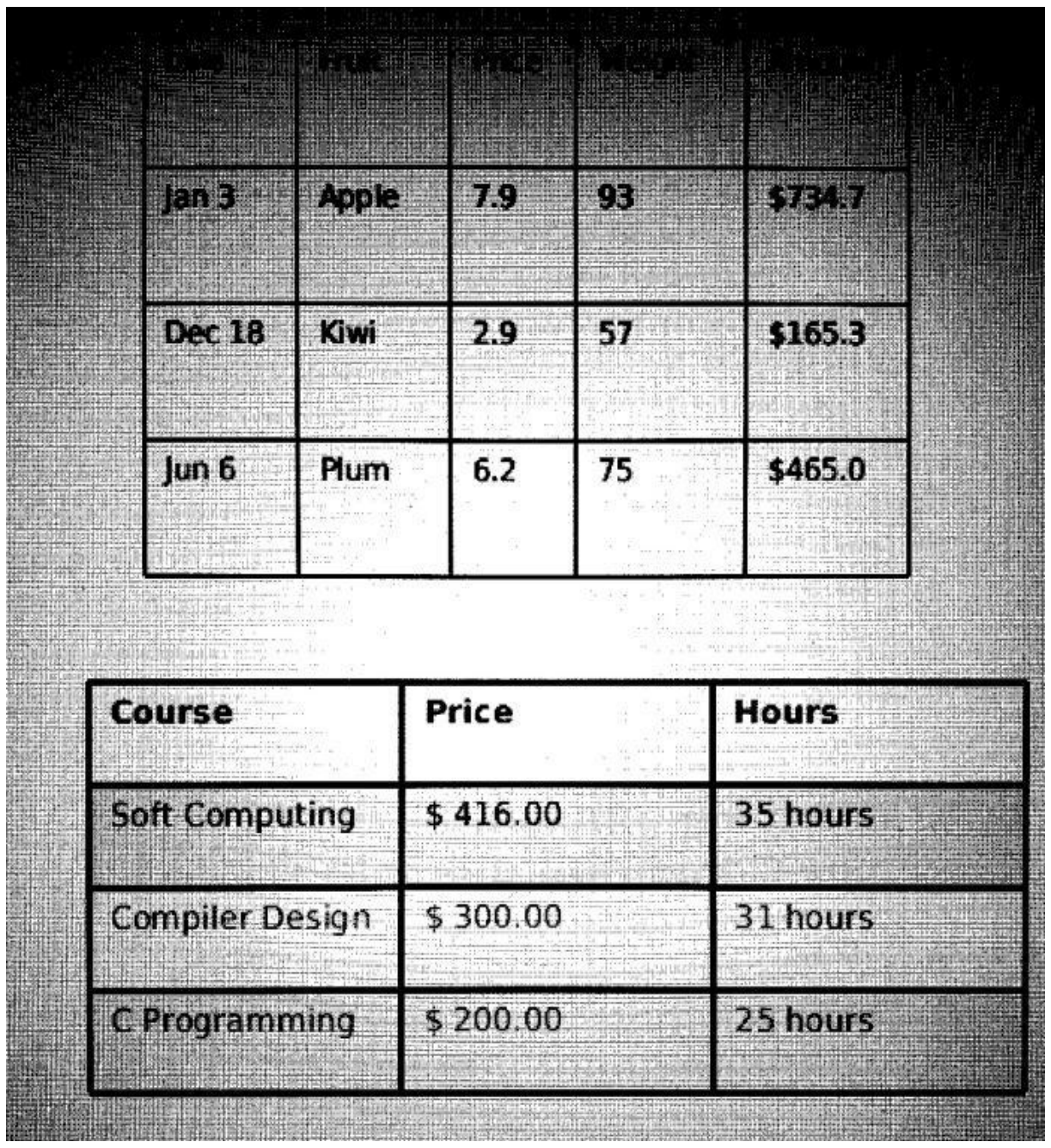

(a) Otsu thresholding on grayscale image

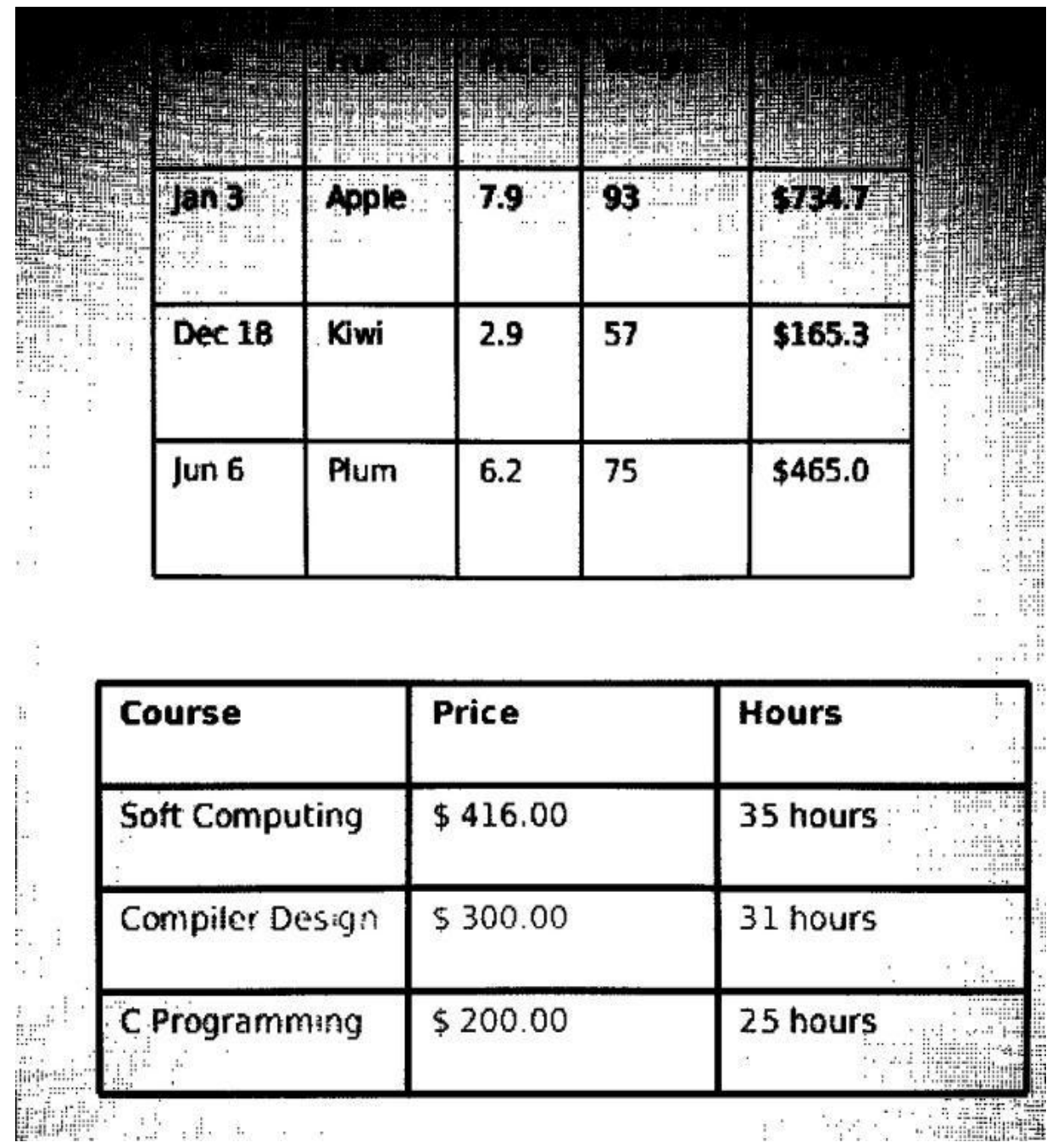

(c) Opening on image a

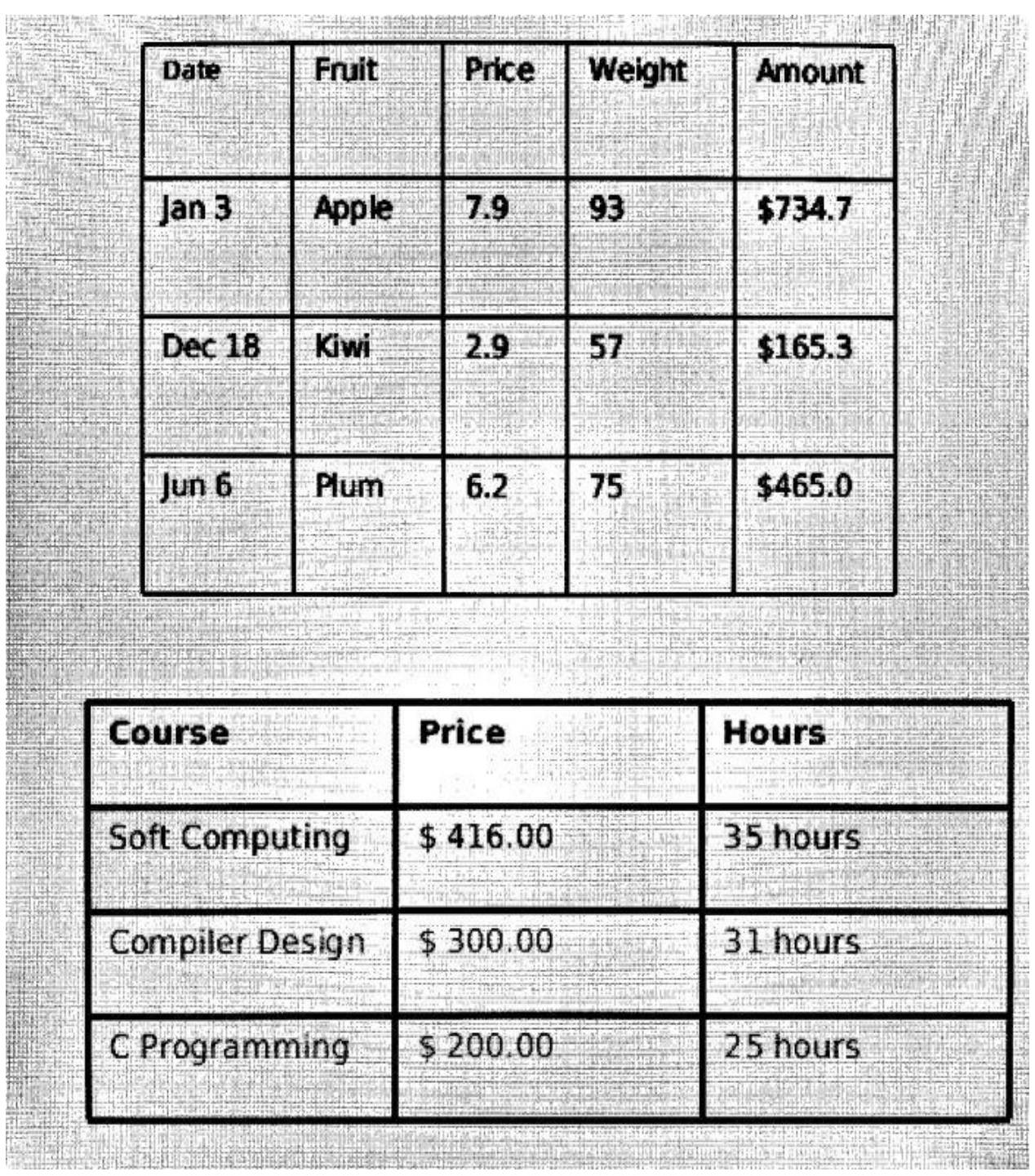

(b) Adaptive Gaussian thresholding on grayscale image

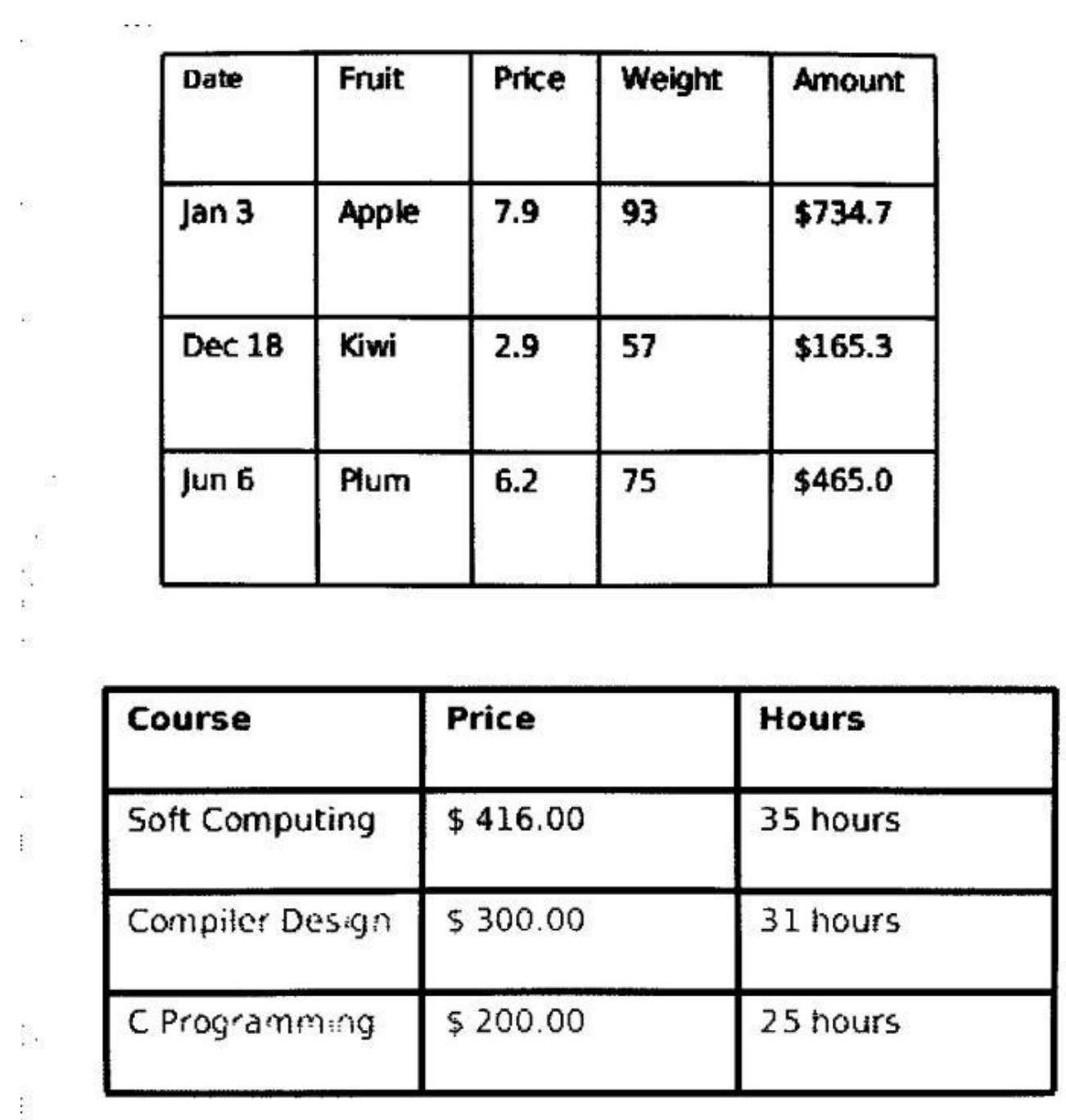

(d) Opening on image b

Fig 5: Applying Otsu's and Adaptive Gaussian thresholding on image and then performing opening which is dilation followed by erosion.

We first apply Otsu binarization with 128 as the threshold and then perform opening to reduce noise. A 2x2 Structuring element performs dilation and erosion twice to produce the result in Fig 5(c). However that result can't be used efficiently because we get darker shades at the top and bottom area of the image. Since the initial image has non-uniform contrast we need to use different values of threshold at different area of image. That is done by Adaptive Gaussian thresholding as in fig 5(b). We get reduced noise after the same opening operation as in Otsu binarization. The block size for Adaptive thresholding is 199 and the value of C is 40. C is just a parameter which is subtracted from the weighted mean calculated.

Next we try to extract the complete table structure from the image. The text and other characters are removed and only vertical and horizontal lines are combined and further processed.

The lines are extracted by morphological operations using horizontal and vertical structuring elements. The kernel length is page height by 80. The complete image is inverted because contours are found for which patches on black background.

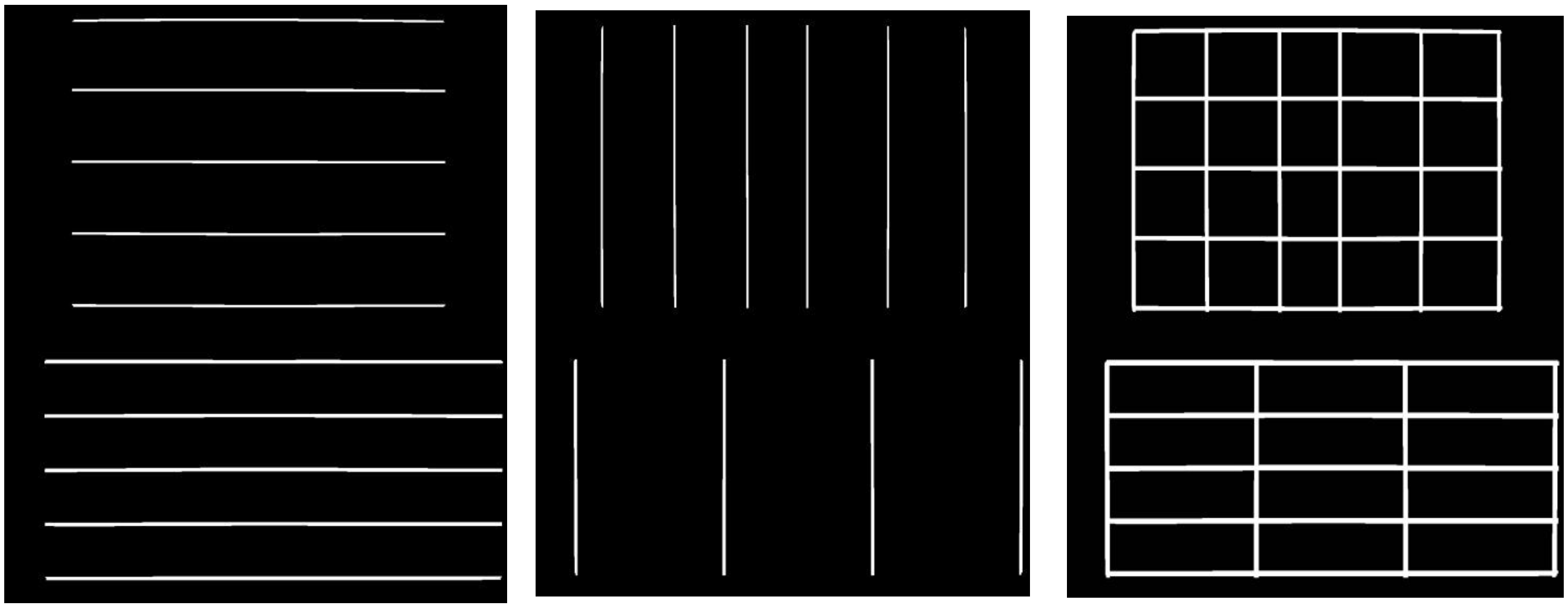

(a) (b) (c)

Fig -6: (a) Horizontal Lines extracted from image (b)Vertical lines extracted from image c combined image of a and b.

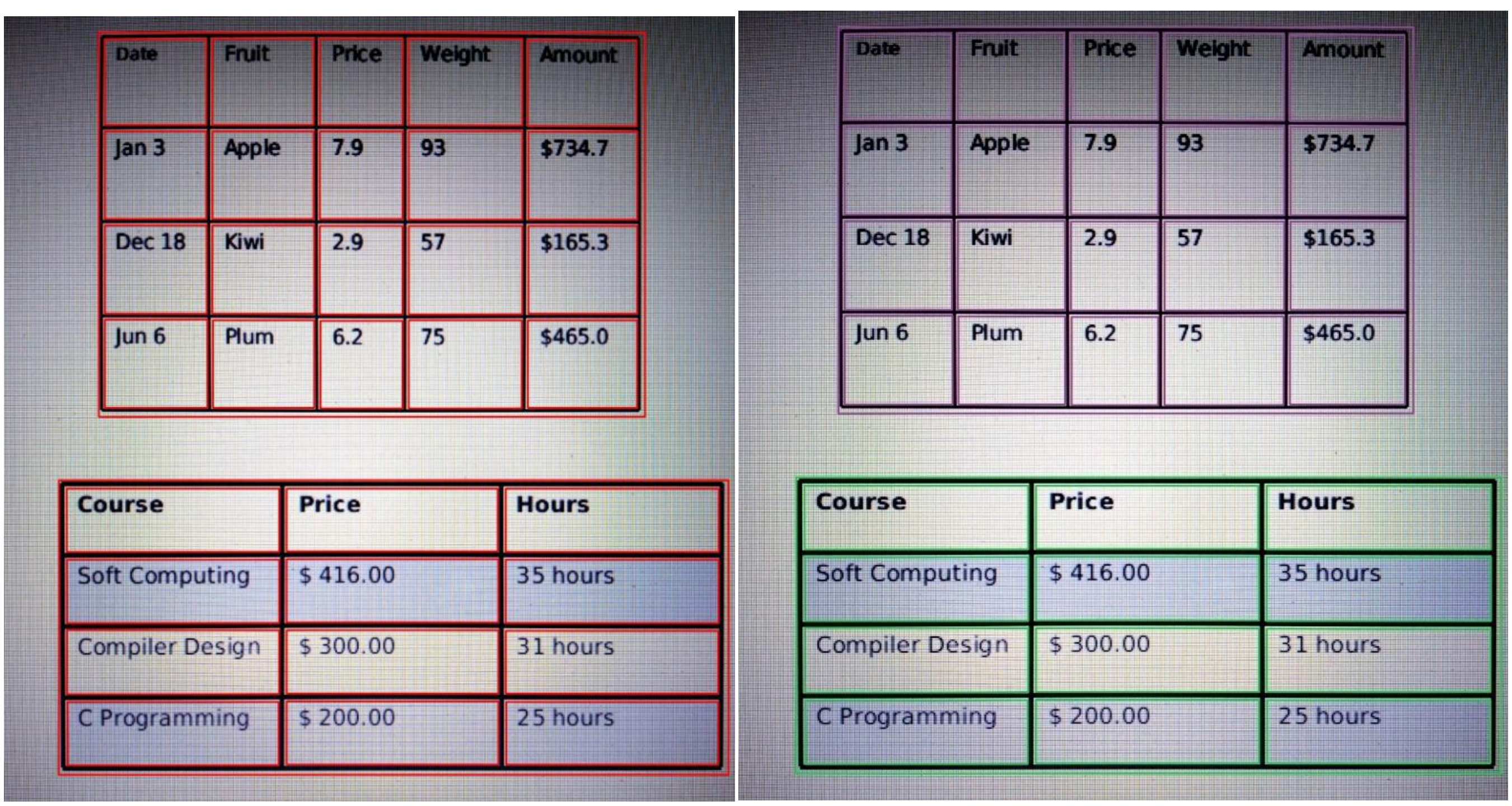

(a) (b)

Fig -7: (a) All contours in the image (b) Grouping and separation of the contours according to tables.

Once the contours are found (Fig 7(a)), they are grouped according to the table they belong(Fig 7(b)). Tesseract OCR engine image_to_data function's output is processed using table initialization algorithm to yield the following output.

| 0 | 1 | 2 | 3 | 4 |
|---|---|---|---|---|
| Date | Fruit | Price | Weight | Amount |
| Jan3 | Apple | 7.9 | 93 | $734.70 |
| 18-Dec | Kiwvi | 2.9 | 57 | $165.30 |
| 6-Jun | Plum | 6.2 | 75 | $465.00 |

| 0 | 1 | 2 |
|---|---|---|
| Course | Price | Hours |
| Soft Computing | $416.00 | 35 hours |
| Compiter Desgn | $300.00 | 31 hours |
| C Programmsng | $200.00 | 25 hours |

Fig -8: CSV image of both the tables extracted

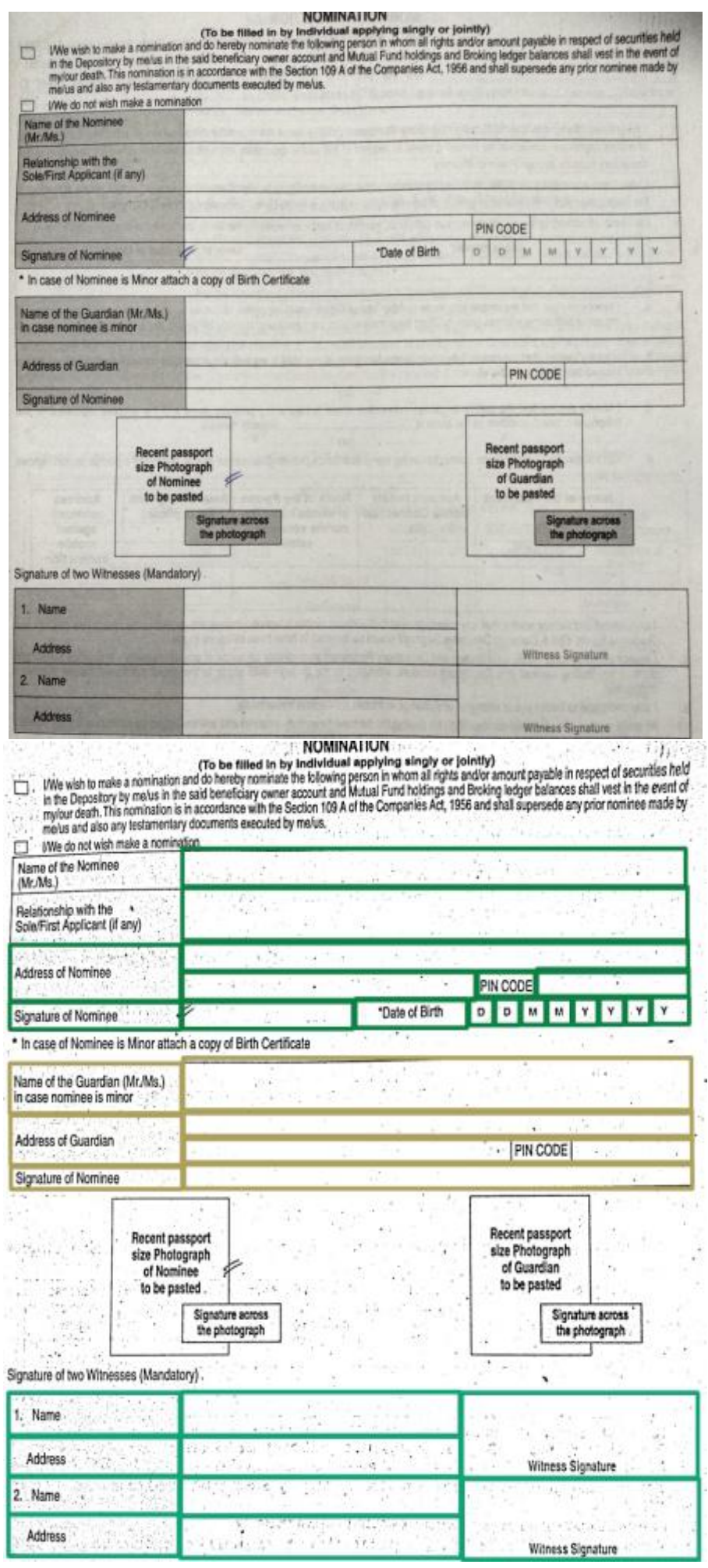

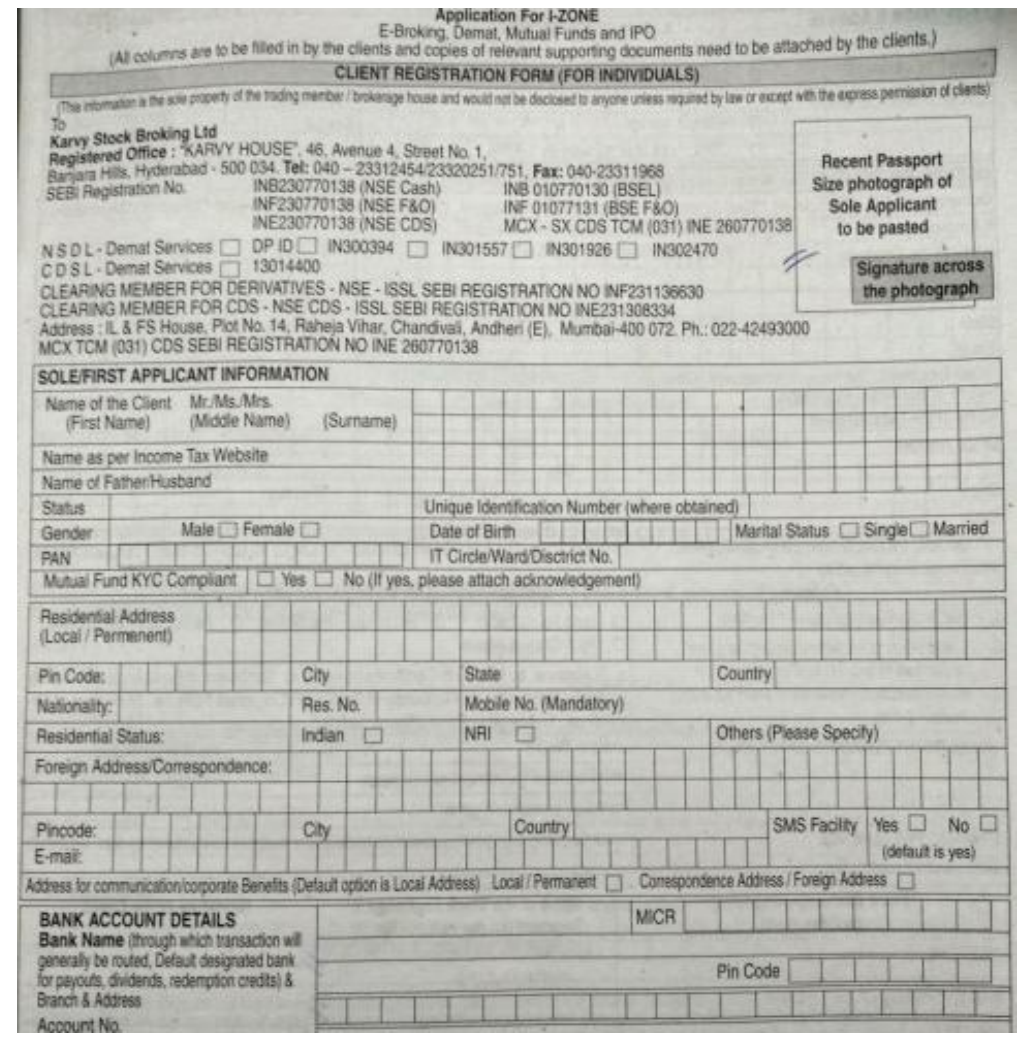

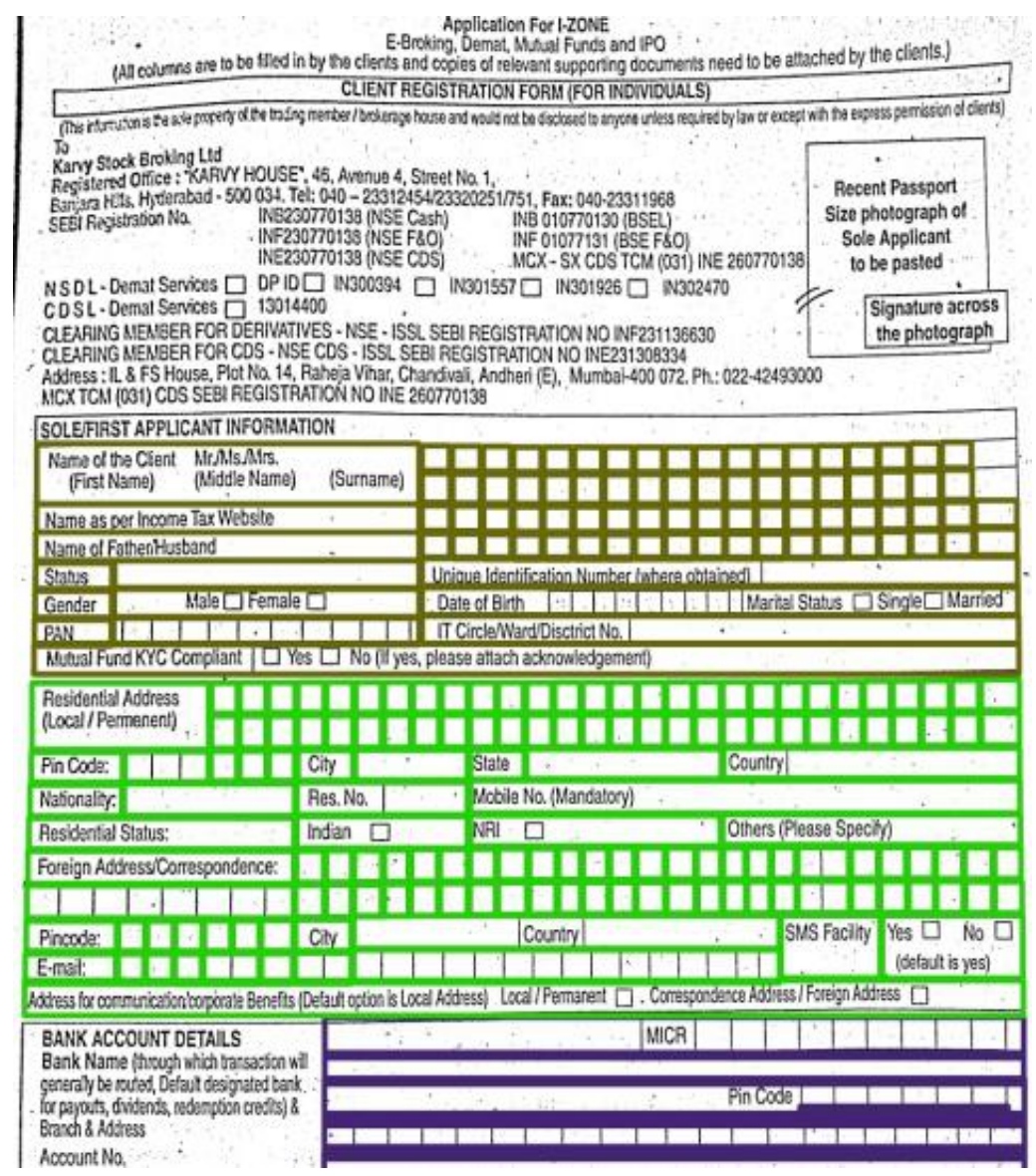

Fig -9: RAW Grayscale images and all table contour detection

## VI. Conclusion

Our stage wise devised algorithm successfully detects and extracts multiple tables from OCR documents or images by clustering them based on their x and y coordinates to derive which corresponding row and column they belong to. This extraction works efficiently on different types of uniformly distributed along with non-uniformly intensity distributed images. The novelty of our algorithm over previous research works is that after successful detection of tables in heterogeneous images, the algorithm also extracts/clones the tables into usable format. It efficiently interprets complex indexing of tables and cases of multiple headers. This algorithm however faces issues in extracting data from borderless tables since the extraction is based on table contours. This opens scope for future researches.